\documentclass[english,aps,prstper,reprint,showpacs,titlepage,longbibliography]{revtex4-1}

\usepackage[T1]{fontenc}
\usepackage[latin9]{inputenc}
\usepackage{geometry}
\usepackage{graphicx}
\usepackage[above,below]{placeins}
\usepackage{times}

\usepackage{hyperref}
\hypersetup{colorlinks=true,urlcolor=blue,citecolor=blue,linkcolor=blue}
\usepackage{enumitem}
\setlist{nosep}

\usepackage{etoolbox}
\AtBeginEnvironment{quote}{\itshape}

\usepackage{textcomp}

\usepackage{censor}

\begin{document}

\begin{titlepage}

  \title{
    Characterizing and monitoring student discomfort in upper-division quantum mechanics
  }

  \author{Giaco Corsiglia}
  \affiliation{\textsuperscript{1}Department of Physics, University of Colorado, Boulder, Boulder, Colorado, 80309, USA}

  \author{
    Tyler Garcia\textsuperscript{2},
    Benjamin P. Schermerhorn\textsuperscript{2,3},
    Gina Passante\textsuperscript{3},
    Homeyra Sadaghiani\textsuperscript{2},
    and Steven Pollock\textsuperscript{1}
  }
  \affiliation{\textsuperscript{2}Department of Physics \& Astronomy, California Polytechnic University Pomona, Pomona, California, 91768, USA}
  \affiliation{\textsuperscript{3}Department of Physics, California State University, Fullerton, Fullerton, California, 92831, USA}

  \begin{abstract}
    \noindent
    We investigate student comfort with the material in an upper-division spins-first quantum mechanics course.  Pre-lecture surveys probing students' comfort were administered weekly, in which students assigned themselves a ``discomfort level'' on a scale of 0--10 and provided a written explanation for their choice. The weekly class-wide average discomfort level was effectively constant over the semester, suggesting that the class found no single unit especially jarring nor especially easy.  Student written responses were coded according to their reported source of discomfort---\textit{math, math-physics connection, physics,} and \textit{notation}.  The relative prevalence of these categories varied significantly over the semester, indicating that students find that different units present different challenges, and also that some of these challenges fade in importance as the semester progresses.  Semi-structured interviews with students in a similar quantum mechanics course at a different institution provided additional context and insight into these results.
    \clearpage
  \end{abstract}

  \maketitle
\end{titlepage}

\StopCensoring

\section{Introduction}

\noindent
Upper-division quantum mechanics (QM) is known to be both conceptually and mathematically challenging for students \cite{Singh2015Review}.  Much research has focused on student difficulties, but recent work has also investigated student perceptions of the course---e.g., examining students' ideas about the nature of quantum mechanics---as a way to guide instruction and future studies \cite{Singh2015Review, Dreyfus2019Splits}.
Johansson discusses students' perceptions of their QM courses in light of cultural expectations, arguing that an expectancy mismatch can affect how students identify as physicists \cite{Johansson2018Undergraduate}.  Schermerhorn et al.\ investigated whether students considered the physics or math more challenging when studying spin systems or spatial wave functions \cite{Schermerhorn2019}.  Outside of the QM context, Gupta et al.\ explored connections between students' emotions and conceptual reasoning \cite{Gupta2018}.

We investigate students' self-reported sense of comfort with the material in QM.   Prompting students to reflect on their own understanding of recently covered material and encouraging them to identify challenging aspects has precedent in the Just-in-Time-Teaching paradigm \cite{Lasry2014}.  In this paper, we present the results of a survey administered on a weekly basis in an upper-division spins-first QM course, as well as complementary preliminary results from an interview study with students from a similar course at a different university.  On the survey, students were asked to identify their level of discomfort with the material along with reasons for that discomfort.  In the interviews, students were prompted to reflect on the content they were learning and their sense of comfort with it.

A spins-first approach is one of two common paradigms for upper-division QM instruction \cite{Sadaghiani2016}.  Where a positions-first approach focuses on solving the Schr\"{o}dinger equation for continuous wave functions that describe the position- or momentum-space distribution of a particle, a spins-first approach instead begins with spin-\textonehalf{} systems described by a discrete, two-state basis before a transition is made to studying continuum systems near the end of the semester.  The spins-first structure emphasizes fundamental quantum mechanical concepts in part by postponing development of the more computationally intensive mathematical formalism required for continuous wave functions \cite{Sadaghiani2016}.  However, one concern is that students may find the discrete-to-continuous transition especially difficult or jarring \cite{DeLeone2017Talk}.

A marked increase in student discomfort following the course's switch to studying spatial wave functions would highlight the aforementioned concern about the difficulty of this transition point.  More generally, the degree of and reasons for student discomfort with the material illuminate the things that students consider important in their own learning, which can inform instruction as well as future research.

\section{Methodology}

\noindent
This paper focuses on student responses to a weekly pre-lecture survey administered in one semester of upper-division QM at University A, \xblackout{a large, public, primarily undergraduate and Hispanic-serving institution}.  Preliminary analysis of these data also prompted an interview study with students in the first semester upper-division QM course at University B, \xblackout{an R1, PhD-granting institution}.  The interview study was developed during and ran concurrently with continued analysis of the survey data.  Both courses adhered to a ``spins-first'' QM curriculum using McIntyre's text~\cite{McIntyre2012}.  \xblackout{The course at University A was taught by physics education research faculty.}

\subsection{Pre-lecture surveys}

\noindent
A pre-lecture assignment was administered online before each class to students in an (in-person) upper-division QM course at University A, where a total of 26 students responded at least once.  Students received participation credit for completing the survey, but it was not graded for correctness.  Beginning in the semester's third week, the first survey of each week opened with the same two questions asking students to reflect on their comfort with the material they were learning.

Students were asked to rank their discomfort level on a scale of 0--10 as follows:

\begin{quote}
  On a scale of 0--10, rate your discomfort (or comfort) with the ideas presented in class this week.

  (We are not asking about how well you think you can answer homework or exam questions, but how the concepts and ideas are ``sitting'' with you and your intuition about the world.)

  0 --- (no discomfort: ``all these ideas make complete sense and seem reasonable'')

  5 --- (moderate discomfort: ``some of the ideas seem illogical and bother me, but I can see what's going on'')

  10 --- (complete discomfort: ``none of these ideas make any sense and I'm deeply concerned'')

  Think of this as a ``quantum pain scale.''
\end{quote}

\noindent Students were then asked to explain their discomfort level:

\begin{quote}
  If you selected a number greater than 0, give a concrete example of an idea or concept that makes you uncomfortable.
\end{quote}

\noindent Not all students provided explanations on every survey, but usually they wrote one or two sentences.

We examined weekly and individual-student averages of the numerical responses.  The written responses were categorized using an emergent coding strategy to identify the type of material the student reported discomfort with (see Sec.~\ref{sec:results-codes}).

Once the coding scheme was agreed upon, codes were assigned to all responses by two researchers working independently, with 75\% initial agreement.  Mismatches were resolved in discussion between the two researchers until 100\% agreement was reached.  Individual responses were assigned multiple codes if the student reported multiple sources of discomfort, meaning that some disagreements were resolved by adopting the codes suggested by both researchers.

\subsection{Semi-structured interviews}

\noindent
In the semester after the pre-lecture survey data were collected, an interview study was performed to probe the reasons for student discomfort in more detail.  Interviewees were all drawn from the first semester upper-division QM course at University B.  The same six volunteers were interviewed individually several times, including during the \textit{time evolution} unit in weeks 6--7, and immediately after the \textit{infinite square well} unit---the first unit in the wave functions context---in week 10. The volunteers were compensated for participation.

The interviews were conducted in spring 2020, and the second round took place one week after University B transitioned to remote learning due to COVID-19.  Instruction on the \textit{infinite square well} unit preceded the switch.  (The pre-lecture surveys were administered in an unaffected semester.)

The interview protocol prompted students to share how they were feeling overall in the course, to assess their comfort level with material in general, and to reflect specifically on their comfort level with certain topics, including notation, math, and physics (i.e., physical concepts).  The protocol also included additional questions outside the scope of this paper.  Analysis of the interview data is ongoing, but preliminary results are presented in Sec.~\ref{sec:discussion}.

\section{Results \label{sec:results}}

\noindent
In this section, we present student responses to both questions on the pre-lecture surveys.  First, we present the average self-reported discomfort level on individual-student and weekly bases.  Then we explore qualitative analysis of students' written explanations for their reported discomfort level.

\subsection{Tracking discomfort level over the semester}

\begin{figure}
  \includegraphics[width=\linewidth]{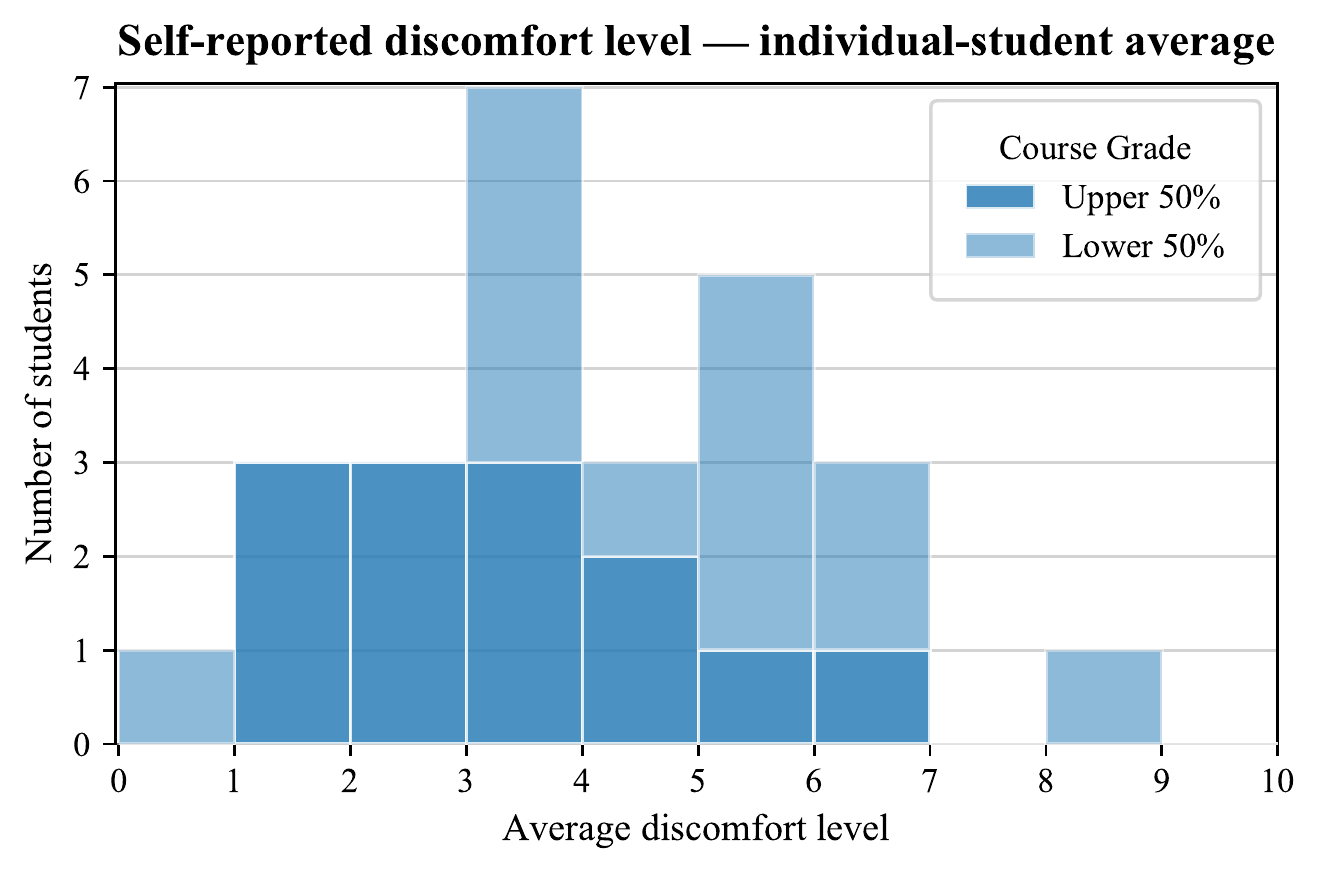}
  \caption{The histogram shows the distribution of students' average self-reported discomfort levels.  Averages are calculated over all of the pre-lecture surveys that the given student responded to.  The individual-student standard deviation averaged over all students is 1.5.  The outlying student on the right only reported their discomfort level on one survey, and the one on the left exclusively reported zeroes.  The darker (lighter) colored bars indicate students whose final course grade fell in the upper (lower) 50\% of the class. \label{fig:student-avg}}
\end{figure}

\noindent
Figure~\ref{fig:student-avg} presents the distribution of students' discomfort levels averaged over the entire semester.  Centered around the overall average of 4, the distribution is reasonably normal.  Individual student responses changed on a week-by-week basis by $\pm1.5$ on average.  The figure shows some correlation between higher discomfort score and poorer course performance, but it is unclear if this reflects students' ability to diagnose their own difficulties, their emotional response to the grades they received on assignments, or another effect.

\begin{figure}
  \includegraphics[width=\linewidth]{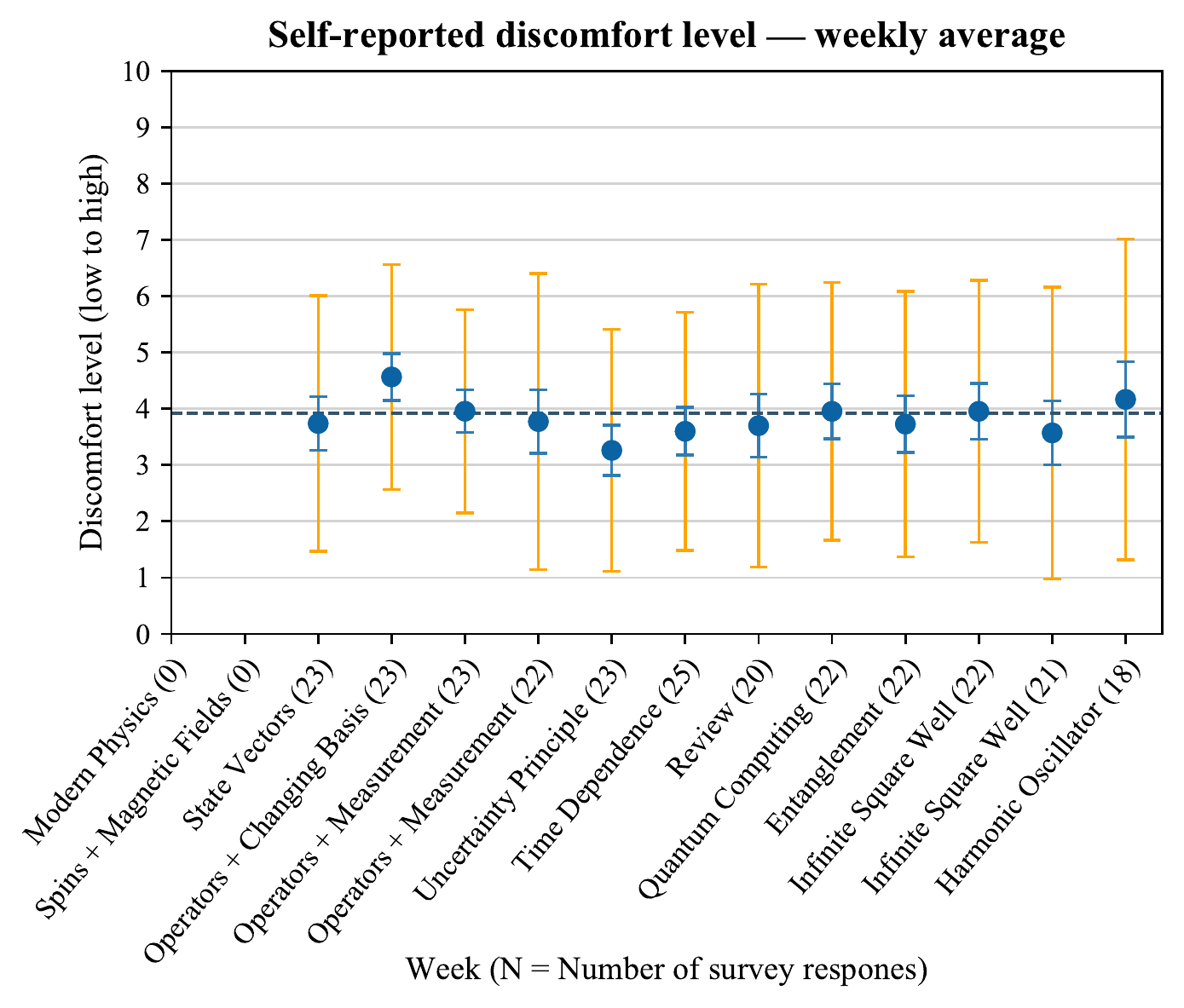}
  \caption{Each dot shows the average discomfort level reported by all students on the pre-lecture survey taken when the most recently covered topic was the one given by the corresponding label on the horizontal axis.  The topics are arranged in the order they were taught, and the labels include the number of students that responded to the corresponding pre-lecture.  The blue (smaller) error bars show the standard error of the mean, and the orange (larger) error bars give the standard deviation. The dashed line shows the overall average. \label{fig:weekly-avg}}
\end{figure}

This quantitative discomfort scale is, at best, calibrated on an individual-student basis.  That is, we hope that a 4$/$10 discomfort level means approximately the same thing to the same student each week, but we cannot assume it means the same thing to every student.  This makes it difficult to compare the magnitude of discomfort level across students.  Still, averaging over students may capture class-wide shifts in discomfort.  Figure~\ref{fig:weekly-avg} shows the class-wide average discomfort level reported on each of the 12 weekly pre-lecture surveys administered over the semester.

The class-wide average discomfort level is effectively constant over the entire semester: it hovers close to 4 and varies by less than 1.  Additionally, there is a sizable spread every week: the standard deviation is usually 2 or higher.  Each week, some students report a high level of discomfort while others report a low level, but there is never a net trend towards higher or lower class-wide discomfort, regardless of changes in the topical areas being covered.

\subsection{Identifying categories of discomfort and tracking their prevalence over the semester \label{sec:results-codes}}

\noindent
Qualitative analysis focused on students' explanations for their discomfort level on the pre-lecture surveys.  Through emergent coding, we identified four categories of discomfort corresponding with different aspects of the material: \textit{math, math-physics connection, physics,} and \textit{notation}.  We also employed a fallback category, \textit{other}.  We define each of these categories as follows.  The percentages indicate the frequency of the given category amongst the 241 total assigned codes.

\textbf{Math} (30\%): The response references either conceptual or procedural understanding of the mathematics, such as discomfort computing integrals.  For example, \textit{``The concepts weren't too difficult I think I just need to practice the calculations \dots{}.''}

\textbf{Math-Physics} (16\%): The response references the connection between the math and the physics, such as discomfort with the physical meaning of an equation. For example, \textit{``I'm mostly uncomfortable with knowing how and when to use the equations.''}

\textbf{Physics} (24\%): The response references physical concepts, such as questions about measurement.  For example, \textit{``I am confused but intrigued by entanglement \dots.''}
This category also includes responses that reference specific content but are too vague for another category, such as, \textit{``I was just having some problems with the clicker questions for measurement and measuring uncertainty \dots{}.''}

\textbf{Notation} (10\%): The response references notation used in the course, usually Dirac notation.  For example, \textit{``The Dirac notation and braket rules are still fuzzy \dots{}.''}

\textbf{Other} (20\%): The response does not reference specific content covered in the course.  Examples include, \textit{``Exams are coming and i am having anxiety \dots,''} and \textit{``Still getting the hang of things.''}  We focus on the four content-centered categories in this work.

\begin{figure}[t]
  \includegraphics[width=\linewidth]{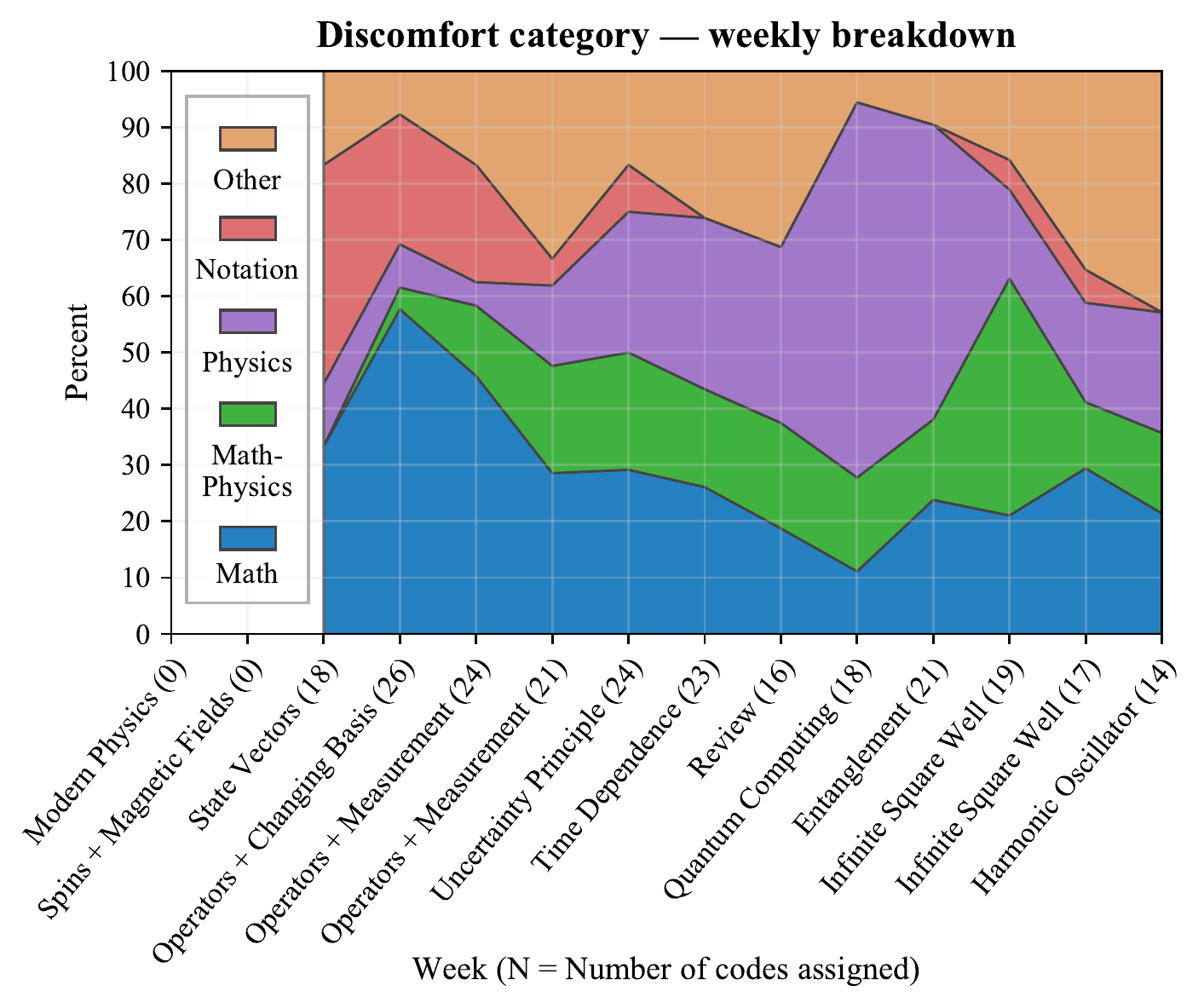}
  \caption{The vertical cross-sections show the distribution of codes assigned to students' written explanations for their self-reported discomfort level on each of the pre-lecture surveys.
  Each cross-section is normalized to the total number of codes assigned to student responses for the given week---this number is included in the labels on the horizontal axis.
  The topics in the labels are the same as in Fig.~\ref{fig:weekly-avg}.
  Some student responses were assigned multiple codes, and not all students provided a written explanation for their reported discomfort level.
   \label{fig:weekly-codes}}
\end{figure}

Overall, more students cite discomfort with \textit{math} than any other category, with \textit{physics} being the second-most common, followed by \textit{math-physics} and \textit{notation}.  However, this ranking was not consistent on a week-by-week basis.  Figure~\ref{fig:weekly-codes} presents the distribution of discomfort categories for each week that the pre-lecture survey was administered.

Early on, \textit{math} and \textit{notation} are the dominant sources of discomfort.  While students report discomfort with \textit{math} throughout the semester, the \textit{notation} category all but disappears by the fourth pre-lecture.  Meanwhile, students report increasing discomfort with physical concepts as the semester progresses, peaking during the units on quantum computing and entanglement, where the \textit{physics} category dominates.

The \textit{physics} category drops back to below 20\% during the ``Infinite Square Well'' unit, which is when the course switches to discussing continuous wave functions as opposed to discrete spin systems.  We also see that \textit{math-physics} is the dominant source of discomfort during this unit, accounting for about 40\% of all reported discomfort, more than twice its maximum in any other week.  We also see a small resurgence of the notation category at this time, with two students referencing notation as a source of discomfort in either this unit or the following one.  Finally, we note that the \textit{other} category surpasses 20\% only in the one or two weeks preceding each exam.

Most individual students' responses included a diversity of categories over the semester.  One student did not provide a single written response.  Of the remaining students, all 25 reported discomfort with \textit{math} at least once, 24$/$25 with \textit{physics}, and 19$/$25 with \textit{math-physics}, although \textit{notation} only arose for 15$/$25 students.  Only 10$/$25 students reported discomfort with any single category more than 50\% of the time (3 of whom favored \textit{other}).

To investigate the relationship between students' discomfort level and categories, we generated a version of Fig.~\ref{fig:weekly-codes} wherein each occurrence of each category was weighted by the discomfort level reported by the given student in the given week.  This did not produce a meaningful change, indicating that there was minimal correlation between discomfort category and discomfort level.

\section{Discussion \& Interview Outcomes \label{sec:discussion}}

\noindent
The transition from discussing discrete spin systems to continuous position-space wave functions did not produce any shift in class-wide discomfort at University A, contrary to our expectations \cite{DeLeone2017Talk}.  Although conducted with a different set of students at a different school, several potential explanations arose in the second round of interviews held at University B, which immediately followed this transition point in a similarly structured QM course.  Discussing the material in the wave functions unit, one student remarked,

\begin{quote}
  It's starting to look a little more familiar to what I've done before because we've started going over, like finite and infinite square wells. We've also started looking at waves which we're also doing in E\&M.
\end{quote}

\noindent
Students have often seen examples of position-space systems like the infinite square well in previous courses, and the mathematics of waves may also feel familiar from other courses.  Another student said of the wave functions unit that, compared with earlier material, she did \textit{``better with the conceptual stuff just because for me it's---it feels more applied.''}  Other interviewees echoed this sentiment: the familiar physical meaning of ``position'' may help students make sense of the new system despite new mathematical formalism.

Based on qualitative data, however, the transition point still warrants instructor attention.
One interviewee remarked, \textit{``I feel like the material got really hard last week, very quickly,''} and others reported that it took them time to find connections between the two halves of the course.  The courses at both universities (A and B) included similar activities at the start of the wave functions unit designed specifically to ease the transition point, and several interviewees at University B referenced this activity as helpful \cite{physport}.

Several patterns arose in the coding of students' written explanations for their reported discomfort level.  First, although Dirac notation is a clear source of discomfort for students when it is introduced at the beginning of the semester, discomfort with notation fades quickly.  Interviewed students shared a similar sentiment, and some expressed that they had come to appreciate the new notation by the end of the semester.
These results suggest that the existing curriculum's approach to this topic worked well for students despite their initial discomfort with Dirac notation.

We also found that more students reported discomfort with the mathematics in the first half of the semester, despite the fact that the mathematics required in the second half tends to be more computationally intensive.  Interviewed students echoed this sentiment.  For example, one student said he considered the class to be \textit{``70\% math, 30\% physical physics understanding''} in his first interview.  He, along with other interviewees, pointed out that they were less familiar with linear algebra and matrix manipulation than with integration and differential equations at the start of the course.

Not all interviewees felt this way, however (see also related research in \cite{Schermerhorn2019}).  One remarked in his first interview that, \textit{``I'm way more confident in my math abilities in this course than I was in like diff-EQ or calc 3,''} although he still considered problem-solving in the course to be \textit{``all math.''}   His perspective shifted in the second interview, saying, \textit{``this class is much more about understanding your approach to the problem than the math itself.''}  This appeared to track the survey results, where units on the uncertainty principle and time dependence saw increases in student discomfort with physical concepts and a decrease in discomfort with the mathematics compared with earlier in the semester.  This trend saw its peak during the units on quantum computing and entanglement, suggesting that those topics were especially effective at drawing out physical concepts.

\section{Conclusion}

\noindent
Weekly surveys probing students' comfort with the material were administered in an upper-division QM course.  Students' average level of discomfort remained constant over the semester, but the reasons for that discomfort varied as the course progressed and as different topics were covered.  We believe our results can guide instructors and researchers towards topical areas that students may feel deserve attention.  We do not, however, consider student discomfort with the course's material inherently bad.  Our results can be seen as identifying the aspects of the material that are top-of-mind for students at different points in their QM class.  For example, the increase in the importance of the \textit{physics} discomfort category during the quantum computing and entanglement units may be an argument for teaching these topics.

Similarly, discomfort with quantum mechanics is common among physicists and could suggest an expert-like understanding of the material.
This paper previews an ongoing analysis of an interview study, which will also examine students' comfort with quantum weirdness.

\acknowledgments{\noindent
\xblackout{We are grateful to the authors' respective PER groups, and to the students in both courses included in our study.  This work has been supported in part by the NSF under Grants No. DUE-1626594, 1626280, and 1626482.}}

\bibliography{bibliography}

\begin{thebibliography}{10}%
\makeatletter
\providecommand \@ifxundefined [1]{%
 \@ifx{#1\undefined}
}%
\providecommand \@ifnum [1]{%
 \ifnum #1\expandafter \@firstoftwo
 \else \expandafter \@secondoftwo
 \fi
}%
\providecommand \@ifx [1]{%
 \ifx #1\expandafter \@firstoftwo
 \else \expandafter \@secondoftwo
 \fi
}%
\providecommand \natexlab [1]{#1}%
\providecommand \enquote  [1]{``#1''}%
\providecommand \bibnamefont  [1]{#1}%
\providecommand \bibfnamefont [1]{#1}%
\providecommand \citenamefont [1]{#1}%
\providecommand \href@noop [0]{\@secondoftwo}%
\providecommand \href [0]{\begingroup \@sanitize@url \@href}%
\providecommand \@href[1]{\@@startlink{#1}\@@href}%
\providecommand \@@href[1]{\endgroup#1\@@endlink}%
\providecommand \@sanitize@url [0]{\catcode `\\12\catcode `\$12\catcode
  `\&12\catcode `\#12\catcode `\^12\catcode `\_12\catcode `\%12\relax}%
\providecommand \@@startlink[1]{}%
\providecommand \@@endlink[0]{}%
\providecommand \url  [0]{\begingroup\@sanitize@url \@url }%
\providecommand \@url [1]{\endgroup\@href {#1}{\urlprefix }}%
\providecommand \urlprefix  [0]{URL }%
\providecommand \Eprint [0]{\href }%
\providecommand \doibase [0]{http://dx.doi.org/}%
\providecommand \selectlanguage [0]{\@gobble}%
\providecommand \bibinfo  [0]{\@secondoftwo}%
\providecommand \bibfield  [0]{\@secondoftwo}%
\providecommand \translation [1]{[#1]}%
\providecommand \BibitemOpen [0]{}%
\providecommand \bibitemStop [0]{}%
\providecommand \bibitemNoStop [0]{.\EOS\space}%
\providecommand \EOS [0]{\spacefactor3000\relax}%
\providecommand \BibitemShut  [1]{\csname bibitem#1\endcsname}%
\let\auto@bib@innerbib\@empty
\bibitem [{\citenamefont {Singh}\ and\ \citenamefont
  {Marshman}(2015)}]{Singh2015Review}%
  \BibitemOpen
  \bibfield  {author} {\bibinfo {author} {\bibfnamefont {Chandralekha}\
  \bibnamefont {Singh}}\ and\ \bibinfo {author} {\bibfnamefont {Emily}\
  \bibnamefont {Marshman}},\ }\bibfield  {title} {\enquote {\bibinfo {title}
  {Review of student difficulties in upper-level quantum mechanics},}\ }\href
  {\doibase 10.1103/physrevstper.11.020117} {\bibfield  {journal} {\bibinfo
  {journal} {Physical Review Special Topics - Physics Education Research}\
  }\textbf {\bibinfo {volume} {11}} (\bibinfo {year} {2015}),\
  10.1103/physrevstper.11.020117}\BibitemShut {NoStop}%
\bibitem [{\citenamefont {Dreyfus}\ \emph {et~al.}(2019)\citenamefont
  {Dreyfus}, \citenamefont {Hoehn}, \citenamefont {Elby}, \citenamefont
  {Finkelstein},\ and\ \citenamefont {Gupta}}]{Dreyfus2019Splits}%
  \BibitemOpen
  \bibfield  {author} {\bibinfo {author} {\bibfnamefont {Benjamin~W.}\
  \bibnamefont {Dreyfus}}, \bibinfo {author} {\bibfnamefont {Jessica~R.}\
  \bibnamefont {Hoehn}}, \bibinfo {author} {\bibfnamefont {Andrew}\
  \bibnamefont {Elby}}, \bibinfo {author} {\bibfnamefont {Noah~D.}\
  \bibnamefont {Finkelstein}}, \ and\ \bibinfo {author} {\bibfnamefont {Ayush}\
  \bibnamefont {Gupta}},\ }\bibfield  {title} {\enquote {\bibinfo {title}
  {Splits in students' beliefs about learning classical and quantum physics},}\
  }\href {\doibase 10.1186/s40594-019-0187-y} {\bibfield  {journal} {\bibinfo
  {journal} {International Journal of STEM Education}\ }\textbf {\bibinfo
  {volume} {6}} (\bibinfo {year} {2019}),\
  10.1186/s40594-019-0187-y}\BibitemShut {NoStop}%
\bibitem [{\citenamefont {Johansson}(2018)}]{Johansson2018Undergraduate}%
  \BibitemOpen
  \bibfield  {author} {\bibinfo {author} {\bibfnamefont {Anders}\ \bibnamefont
  {Johansson}},\ }\bibfield  {title} {\enquote {\bibinfo {title} {Undergraduate
  quantum mechanics: lost opportunities for engaging motivated students?}}\
  }\href {\doibase 10.1088/1361-6404/aa9b42} {\bibfield  {journal} {\bibinfo
  {journal} {European Journal of Physics}\ }\textbf {\bibinfo {volume} {39}},\
  \bibinfo {pages} {025705} (\bibinfo {year} {2018})}\BibitemShut {NoStop}%
\bibitem [{\citenamefont {Schermerhorn}\ \emph {et~al.}(2019)\citenamefont
  {Schermerhorn}, \citenamefont {Villasenor}, \citenamefont {Agunos},
  \citenamefont {Sadaghiani}, \citenamefont {Passante},\ and\ \citenamefont
  {Pollock}}]{Schermerhorn2019}%
  \BibitemOpen
  \bibfield  {author} {\bibinfo {author} {\bibfnamefont {Benjamin~P.}\
  \bibnamefont {Schermerhorn}}, \bibinfo {author} {\bibfnamefont {Armando}\
  \bibnamefont {Villasenor}}, \bibinfo {author} {\bibfnamefont {Darwin~Del}\
  \bibnamefont {Agunos}}, \bibinfo {author} {\bibfnamefont {Homeyra}\
  \bibnamefont {Sadaghiani}}, \bibinfo {author} {\bibfnamefont {Gina}\
  \bibnamefont {Passante}}, \ and\ \bibinfo {author} {\bibfnamefont {Steven}\
  \bibnamefont {Pollock}},\ }\bibfield  {title} {\enquote {\bibinfo {title}
  {Student perceptions of math-physics interactions throughout spins-first
  quantum mechanics},}\ }in\ \href {\doibase 10.1119/perc.2019.pr.Schermerhorn}
  {\emph {\bibinfo {booktitle} {2019 Physics Education Research Conference
  Proceedings}}}\ (\bibinfo {address} {Provo, UT},\ \bibinfo {year}
  {2019})\BibitemShut {NoStop}%
\bibitem [{\citenamefont {Gupta}\ \emph {et~al.}(2018)\citenamefont {Gupta},
  \citenamefont {Elby},\ and\ \citenamefont {Danielak}}]{Gupta2018}%
  \BibitemOpen
  \bibfield  {author} {\bibinfo {author} {\bibfnamefont {Ayush}\ \bibnamefont
  {Gupta}}, \bibinfo {author} {\bibfnamefont {Andrew}\ \bibnamefont {Elby}}, \
  and\ \bibinfo {author} {\bibfnamefont {Brian~A.}\ \bibnamefont {Danielak}},\
  }\bibfield  {title} {\enquote {\bibinfo {title} {Exploring the entanglement
  of personal epistemologies and emotions in students' thinking},}\ }\href
  {\doibase 10.1103/physrevphyseducres.14.010129} {\bibfield  {journal}
  {\bibinfo  {journal} {Physical Review Physics Education Research}\ }\textbf
  {\bibinfo {volume} {14}} (\bibinfo {year} {2018}),\
  10.1103/physrevphyseducres.14.010129}\BibitemShut {NoStop}%
\bibitem [{\citenamefont {Lasry}\ \emph {et~al.}(2014)\citenamefont {Lasry},
  \citenamefont {Dugdale},\ and\ \citenamefont {Charles}}]{Lasry2014}%
  \BibitemOpen
  \bibfield  {author} {\bibinfo {author} {\bibfnamefont {Nathaniel}\
  \bibnamefont {Lasry}}, \bibinfo {author} {\bibfnamefont {Michael}\
  \bibnamefont {Dugdale}}, \ and\ \bibinfo {author} {\bibfnamefont {Elizabeth}\
  \bibnamefont {Charles}},\ }\bibfield  {title} {\enquote {\bibinfo {title}
  {Just in time to flip your classroom},}\ }\href {\doibase 10.1119/1.4849151}
  {\bibfield  {journal} {\bibinfo  {journal} {The Physics Teacher}\ }\textbf
  {\bibinfo {volume} {52}},\ \bibinfo {pages} {34--37} (\bibinfo {year}
  {2014})}\BibitemShut {NoStop}%
\bibitem [{\citenamefont {Sadaghiani}(2016)}]{Sadaghiani2016}%
  \BibitemOpen
  \bibfield  {author} {\bibinfo {author} {\bibfnamefont {Homeyra}\ \bibnamefont
  {Sadaghiani}},\ }\bibfield  {title} {\enquote {\bibinfo {title} {Spin first
  vs. position first instructional approaches to teaching introductory quantum
  mechanics},}\ }in\ \href
  {https://www.compadre.org/per/perc/2016/Detail.cfm?id=6660} {\emph {\bibinfo
  {booktitle} {2016 Physics Education Research Conference Proceedings}}}\
  (\bibinfo {address} {Sacramento, CA},\ \bibinfo {year} {2016})\BibitemShut
  {NoStop}%
\bibitem [{\citenamefont {DeLeone}(2017)}]{DeLeone2017Talk}%
  \BibitemOpen
  \bibfield  {author} {\bibinfo {author} {\bibfnamefont {Charles~Joseph}\
  \bibnamefont {DeLeone}},\ }\href
  {https://www.aapt.org/AbstractSearch/FullAbstract.cfm?KeyID=25825} {\enquote
  {\bibinfo {title} {Bridging the discrete to the continuum gap in quantum
  mechanics},}\ } (\bibinfo {year} {2017})\BibitemShut {NoStop}%
\bibitem [{\citenamefont {McIntyre}(2012)}]{McIntyre2012}%
  \BibitemOpen
  \bibfield  {author} {\bibinfo {author} {\bibfnamefont {David~H}\ \bibnamefont
  {McIntyre}},\ }\href@noop {} {\emph {\bibinfo {title} {Quantum Mechanics: A
  Paradigms Approach}}}\ (\bibinfo  {publisher} {Pearson Education, Inc.},\
  \bibinfo {address} {San Francisco, CA},\ \bibinfo {year} {2012})\BibitemShut
  {NoStop}%
\bibitem [{phy()}]{physport}%
  \BibitemOpen
  \href {https://www.physport.org/curricula/ACEQM/} {\enquote {\bibinfo {title}
  {Adaptable curricular exercises for quantum mechanics:
  https://www.physport.org/curricula/aceqm/},}\ }\BibitemShut {NoStop}%
\end{thebibliography}%

\end{document}